\documentclass[amsmath,amssymb,preprintnumbers, nofootinbib]{revtex4}
\usepackage{latexsym}
\usepackage{epsfig}
\usepackage{amssymb}
\usepackage{amsmath}
\usepackage{color}

\newcommand{\bea}{\begin{eqnarray}}
\newcommand{\eea}{\end{eqnarray}}
\newcommand{\la}{\langle}
\newcommand{\ra}{\rangle}
\newcommand{\etal}{\textit{et al.}}

\begin{document}

\preprint{Draft:Please do not distribute}
\title{Measurement of Flux Fluctuations in Diffusion in the Small-Numbers Limit}
\author{Effrosyni Seitaridou}
\author{Mandar M. Inamdar}
\author{Rob Phillips$^{\rm 1}$}
\email{phillips@pboc.caltech.edu}
\affiliation{Division of Engineering and Applied Science and
$^{\rm 1}$ Kavli Nanoscience Institute, California Institute of Technology,
Pasadena, CA 91125}
\author{Kingshuk Ghosh}
\author{Ken Dill}
\affiliation{Department of Biophysics, University of California, San
  Francisco, CA}
\date{\today}

\begin{abstract}

Using a microfluidics device filled with a colloidal suspension of
microspheres, we test the laws of diffusion in the limit of small
particle numbers.  Our focus is not just on average properties
such as the mean flux, but rather on the features of the entire
distribution of allowed microscopic trajectories that are possible
during diffusive dynamics. The experiments show that: (1) the flux
distribution is Gaussian; (2) Fick's Law
--- that the average flux is proportional to the particle gradient
--- holds even for particle gradients down to one or zero
particles; (3) the variance in the flux is proportional to the sum
of the particle numbers; and (4) there are backwards flows, where
particles flow up a concentration gradient, rather than down it,
and their numbers are well-predicted by theory and consistent with
a new Flux Fluctuation Theorem.
\end{abstract}


\maketitle

\section{Introduction}

Fick's Law, which describes the diffusion of atoms, molecules, and
particles, is important in many areas of science, and is the basis
for engineering models of material transport.  One statement of
Fick's First Law is that the average particle flux is proportional
to the average concentration gradient~\cite{dill03},
\bea  \la J \ra = -D{\frac{\partial \la c \ra} { \partial x}},
\eea
where $\la J \ra$ is the observed macroscopic flux and $\la c \ra$
is the concentration of particles. We use brackets here,
$\la\ldots\ra$, to make explicit that this phenomenological
expression deals with averages over macroscopically large numbers
of particles, and to indicate that the particle concentration and
flux can be meaningfully represented as smooth functions of space
and time in macroscopic systems.  Fick's First Law is the basis
for Fick's Second Law, also known as the diffusion equation,
\bea \frac{\partial \la c \ra} { \partial t} = D \frac{\partial^2
\la c \ra}  {\partial x^2} . \eea
These equations have been extensively verified in bulk gases and
solutions with macroscopically large number of
particles~\cite{baker87}.

Our particular interest here is in the ``small numbers'' limit of
Fick's Law, where there are only a few particles in the system,
with special reference to the fluctuations that attend diffusive
dynamics. Small particle numbers and their fluctuations are
important in nanotechnology; inside biological cells, where the
typical copy number of any given type of protein is often less
than a few thousand~\cite{wu05}; and in single-molecule studies of
ion channels, molecular motors, and in laser trap experiments
~\cite{hille01, howard01, neuman04}. Fick's Law describes averages
over a macroscopic number of particles; it does not describe
small-number fluctuational quantities, such as $\la J^2 \ra - \la
J \ra^2$, or any other aspect of the flux distribution function.
One of our motivations for undertaking this work is growing
theoretical and experimental interest in nonequilibrium dynamics
which centers on the distributions of microtrajectories available
to such systems. We reasoned that a first step in examining the
distribution of microtrajectories in nonequilibrium systems would
be to revisit well-established problems such as diffusion, but
with an eye to explicitly measuring (and calculating) the entire
distribution of microscopic trajectories.

Does Fick's Law hold in the limit of small numbers of particles?
And, are there violations? That is, if Fick's Law predicts flow to
the right, due to a concentration gradient sloping downward
towards the right, does it ever happen that particles flow instead
to the left?  Such situations have been called ``Second-Law
violations''~\cite{evans1,evans02}; or, in classical thermal
problems, they are expressed in terms of ``Maxwell's
Demon''~\cite{feynman96}.  Such fluctuations are, of course, not
real violations of the Second Law, because the Second Law is only
a statement about averages, not fluctuations~\cite{callen85}. In
this article we refer instead to such trajectories that go
``against the grain'' as {\it bad-actors.}  Our interest here is
not just in average fluxes, but in the full flux distribution
function.  Traditionally, diffusion has been studied in the bulk,
where the number of particles is large.  Only more recently has it
become possible to perform experiments on small-numbers diffusion
and on dynamical distribution functions, which explicitly
emphasize the character of the microscopic trajectories that
describe the dynamics of nonequilibrium systems, based on advances
in nanotechnology, video microscopy and microfluidics.

To predict dynamical distributions of diffusion rates, we can use
either classical random-flight theory or a recent
maximum-entropy-like approach~\cite{ghosh06}, called maximum
caliber, based on work of ET Jaynes~\cite{jaynes80}.  In short, if
particles are independent, diffusing in one dimension, and if
their jump rates are stationary in time, the distribution of
particle fluxes, $P(J)$, at time $t$ along an $x$-axis from one
bin at $x$ having $N_1$ particles, to an adjacent bin at $x +
\Delta x$ having $N_2$ particles, should follow the binomial
distribution, or approximately a Gaussian function~\cite{ghosh06},
\begin{eqnarray}
\displaystyle{P(J)}& =\displaystyle{ \frac{1}{{\sqrt{2 \pi \langle (\Delta J)^2\rangle}}}
\exp\left( -\frac{(J - \langle J \rangle)^2}{2 \langle (\Delta J)^2 \rangle}
\right)  } \nonumber \\
&\displaystyle{= \frac{1}{\sqrt{2\pi (pqN)}} \exp\left(-
\frac{(J - p\Delta N)^2}{2pqN}\right), }
\label{eq:fluxdist}
\end{eqnarray}
where $\Delta N = N_1 - N_2$, $N = N_1 + N_2$, $\la \Delta J^2
\ra$ is the variance in the flux $J$, and $q = 1-p$, with $p$
being the probability that a particle jumps in the time interval
$\Delta t$.

Various moments of the distribution function are readily obtained
from this model.  First, the model predicts that the average net
number of particles, $J$, that jump per unit time at time $t$
is~\cite{ghosh06}
\bea \label{eq:J} \la J \ra = \la j_1 - j_2 \ra = -\frac{p}
{\Delta t}\Delta N, \eea
where $j_1$ is the flux from the bin $1$ at $x$ to bin $2$ at
$x+\Delta x$, and $j_2$ is from bin $2$ to $1$. This
proportionality of the average flux $\la J \ra$ to $\Delta N$,
simply predicts Fick's Law, where the diffusion coefficient D is
related to $p$ by $D=p\Delta x^2/\Delta t$, and where $\Delta x$
is the bin size and $\Delta t$ is the unit time step.

For the flux fluctuations, i.e., the second moment, the model
predicts
\bea \label{eq:DJsqr} \la \Delta J^2 \ra =  \la (J - \la J
\ra)^2\ra = \frac{p(1-p)} {\Delta t^2}N, \eea
where $N = N_1 + N_2$ is the total number of particles associated
with the two bins of interest. Hence, the key prediction here is
that the flux fluctuations are proportional to the total particle
number, $N$.

We are also interested in the number of {\it bad-actors,} i.e., the
number of trajectories that would lead to particle flows up a
concentration gradient, rather than down it.  This quantity can be
derived from the flux distribution~\cite{ghosh06} as
\begin{eqnarray}
\Phi_{\rm badactors} &=& \displaystyle{ \frac{1}{ 2}\left(1 -
\textrm{erf}\left( \frac{\la J \ra} {\sqrt{2 \la (\Delta J)^2 \ra}}\right)\right)} \nonumber \\
&\approx& \frac{1} {2} - \frac{1} {\sqrt{2\pi}} \frac{\la J \ra }{
\sqrt{\la
      (\Delta J)^2 \ra}} \ldots \nonumber \\
&+&  {\cal O}\left\{\left(\frac{\la J \ra} { \sqrt{\la (\Delta
J)^2 \ra}}\right)^3\right\}, \label{eq:badactors}
\end{eqnarray}
where the approximation holds for small values of $\la J
\ra/\sqrt{\la (\Delta J)^2 \ra}$.  In the expression above, the
next higher term (the cubic term), is an order of magnitude
smaller than the linear term for the values of $\la J
\ra/\sqrt{\la (\Delta J)^2 \ra}$ used in our experiments (see
Figure~\ref{fig:badactors}).
\subsection{A ``Flux'' Fluctuation Theorem}
Recently, a useful description of nonequilibrium dynamics has
involved fluctuation theorems.  Fluctuation theorems characterize
the extent to which the system deviates from its dominant flow
behavior~\cite{evans02, crookspre, seifert06, bustamante,
feitosa04}. In the diffusive dynamics case of interest here, if
the number of particles $N_1$ in bin $1$ is greater than the
number of particles $N_2$ in bin $2$, then particles, on average,
will flow from $1$ to $2$. Fluctuation theorems describe the
amount of reverse flow. Ours is a {\it flux} fluctuation theorem,
i.e., it is expressed in terms of the quantity $P(J)/P(-J)$, where
$J$ is the flux. This differs from fluctuation
theorems~\cite{crookspre, evans02} expressed in terms of
entropies, $P(\Delta S)/P(-\Delta S)$ and from the work theorems
of Jarzynski and Crooks, which are expressed in terms of the work
$w$ as $P[w/(kT)]/P[-w/(kT)]$~\cite{crooks99}.
In our approach, the ratio $P(J)/P(-J)$, obtained from
Eq.~\ref{eq:fluxdist}, gives the ratio of probabilities of fluxes
in the forward and backward directions~\cite{ghosh06},
\bea \label{eq:DF_fluctthm} \ln {\frac{P(J)} {P(-J)}} = \frac{2\la
J \ra} {\la (\Delta J)^2 \ra}J. \eea
Thus, the quantity $\ln [P(J)/P(-J)]$ is predicted to be
proportional to the normalized flux $\la J \ra/\la (\Delta J)^2
\ra\times J$. In situations having large flux, the back-flow
becomes exponentially negligible. We subjected these predictions
to experimental tests.

\section{The Microfluidics Experiments}
\begin{figure}
\centering
\includegraphics[width = 8cm]{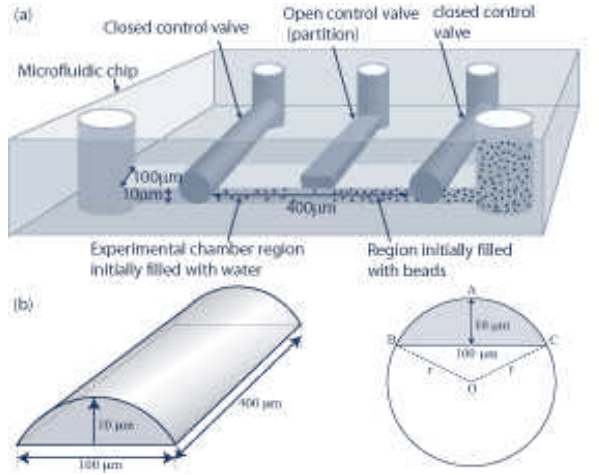}
\caption{\label{fig:setup}The microfluidics experiment. Colloids
corralled on one side of a gate begin to diffuse at time $t=0$ by
opening the gate.  (a) Schematic of the microfluidic chip (see
text for details). (b) The geometry of the microfluidic chamber
(not drawn to scale).}
\end{figure}

To study the dynamical distributions in diffusion, we devised a
microfluidics experiment. Using the techniques of soft
lithography, chip fabrication~\cite{unger00} and the Sylgard $184$
Silicone elastomer kit (Dow Corning Corporation), we made a
microfluidics chamber having approximate dimensions $400\rm{\mu
m}$ by $100\rm{\mu m}$, partitioned into two regions (see
Fig.~\ref{fig:setup}a). The cross-section of this chamber is a
segment of a circular disc, with a maximum depth of $10 \rm{\mu
m}$ (see Fig.~\ref{fig:setup}b). The chamber is filled on one side
with a solution containing about $200$ colloidal, green
fluorescent polystyrene particles $0.29 \rm{\mu m}$ in diameter
(Duke Scientific, Cat. No. G300) (see Fig.~\ref{fig:setup}a). The
beads are at an optimized concentration so that the interactions
are negligible~\cite{refAA} while at the same time permitting
sufficient statistics over a wider range of $\Delta N$ and $N$.

At time $t=0$, we open a microfluidic gate, allowing particles to
diffuse from one side to the other, and we begin taking periodic
snapshots under an Olympus IX71 inverted microscope. (We performed
the same experiment under equilibrium conditions where the initial
concentration was uniform across the whole chamber (results not
shown, see~\cite{refCC})). We take $3$ snapshots of the beads in
the chamber every time interval of $\Delta t = 10$ seconds for $6$
hours. Since there is a possibility that within one snapshot some
particles are temporarily overlapping and/or are out of focus, the
$3$ snapshots are used to minimize that error. The snapshots are
taken using fluorescence microscopy with a SONY DFW-V500 camera.
(During the time when no snapshots are taken, a shutter prevents
the experimental chamber from being exposed to the incident light,
to prevent photobleaching and heating the chamber.) We then
determine the particle positions at each snapshot using a
computerized centroid tracking algorithm~\cite{grier96}. The
time-dependent particle density is determined by dividing the
chamber into a number of equal-sized bins of value $\Delta x$ each
along the longest dimension of $400\rm{\mu m}$ and by computing
the number of particles in each bin as a function of time.
Although the microfluidic chamber is three-dimensional, it can be
shown that, in the case of weak particle-particle and
particle-wall interactions, the problem can be collapsed to a
one-dimension diffusion problem. Therefore, we bin only along the
x-axis, the direction of the concentration gradient. The choice of
the bin size will affect the statistics for each combination of
$N_1$ and $N_2$ as well as the range of $N$ and $\Delta N$
themselves.  If the bin size is too big, then there will not be
enough statistics and, in addition, the range of values for $N$
will not include small numbers (since it will be rare to have $1$
or $2$ particles in a single bin). On the other hand, if the bin
size is too small, we may not have a sufficient range of values
for $N$ and $\Delta N$ (since it will be rare to have more than a
couple of particles in the small bin). Also, in this case, there
will be an increased probability for a particle to have multiple
jumps across bins within the time interval $\Delta t$. Therefore,
the optimal choice of the bin size was made based on the bead's
expected mean excursion within the time-interval $\Delta t$, which
is $\sqrt{2D\Delta t}$. This is the only relevant microscopic
length scale. Here, $D$ is the diffusion coefficient for an
individual bead given by the Stokes's formula~\cite{rief65}. For a
bead of $0.29 \rm{\mu m}$ in diameter suspended in water at room
temperature, the Stokes's formula gives a diffusion coefficient
$D$ of approximately $1.5\rm{\mu m^{2}/ \rm{second}}$. This value,
within experimental error, is equal to the one we obtain by
fitting our data of the concentration profile at different times
to the one dimensional diffusion equation using $D$ as our fitting
parameter (i.e. $D=1.3\pm 0.27\rm{\mu m^{2}/\rm{second}}$). This
gives a bin size of $\Delta x \approx 5\rm{\mu m}$. By observing
all the consecutive bin pairs for all the frames taken we were
able to obtain, on average, about $5000$ points for each
combination of $N_1$ and $N_2$. Given the bead concentration in
the microfluidics channel, the $N_1$ and $N_2$ ranged from $0$
through $6$. The choice of bin size determines the value of the
jump probability $p$, as discussed in~\cite{refDD}.

We can find the flux at a plane $i$ at a specific time interval
from the computed particle distribution statistics as a function
of position $x$ and time $t$ mentioned above. Since the
microfluidic chamber is isolated, the total number of particles
stays the same from one frame to the next.  As a result of this
conservation in particle number, the flux at plane $i+1$
($J_{i+1}$), i.e., the plane that separates bins $i$ and $i+1$,
can be easily evaluated by using the continuity equation,
\begin{eqnarray} \label{eq:deltan}
N_{i} \left( t+\Delta t \right) &=& N_{i} \left(t \right)+ \left(
J_{i} \left( t \right)-J_{i+1} \left( t \right) \right)\Delta t\\
\label{eq:conteq} \Rightarrow J_{i+1}\left( t
\right)&=&-\frac{N_{i} \left( t+\Delta t \right)-N_{i} \left( t
\right)}{\Delta t}+J_{i} \left( t \right),
\end{eqnarray}
where $N_{i}$ is the number of particles in bin $i$. Since the
microfluidic chamber is isolated, from our boundary conditions the
flux $J_0$ (flux at $x = 0$) is zero at all times. Combined with
Eq.~\ref{eq:conteq}, we obtain $J_1(t)$. Thus, from the analysis
of these images, we obtain complete sets of the values of
$\{N_{i}(t)\}$ and $\{J_{i}(t)\}$ in all the bins and at all times
of observation. Then, for each pair of consecutive bins with
specific values of $N_1$ and $N_2$, we construct the histogram of
$J$ values. Upon normalization the histogram becomes the flux
probability distribution, $P(J)$.
\section{Results}

\subsection{The Flux Distribution Function is Gaussian.}
\begin{figure}
\centering
\includegraphics*[width = 8cm]{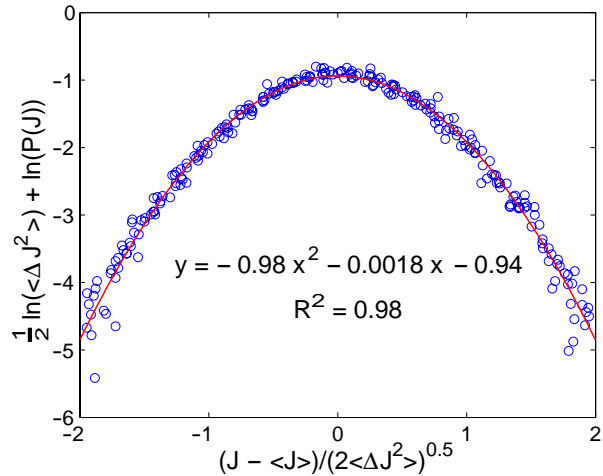}
\caption{\label{fig:mastergaussian} The flux distribution
function. $\frac{1}{2}\ln({\langle (\Delta J)^2 \rangle})+\ln
(P(J))$ is plotted against $(J - \langle J
\rangle)/{\sqrt{2\langle (\Delta J)^2 \rangle}} $, based on the
form indicated by Equation~\ref{eq:fluxdist}.  Circles indicate
experimental points, the line shows a quadratic fit to the data.
The coefficient of determination ($R^2$) for the fit is also
reported. This demonstrates that the distribution function is
Gaussian, and we find that the coefficients are well predicted by
Equation~\ref{eq:fluxdist}.}
\end{figure}

Figure~\ref{fig:mastergaussian} shows our observed particle flux
distribution function at the optimized concentration. All the data
falls on a single master curve where $\langle J \rangle$ and
$\langle (\Delta J)^2 \rangle$ have been calculated separately
from each combination of $N_1$ and $N_2$. The quadratic form
observed on this log plot shows that the distribution function is
Gaussian. The theory predicts that: (i) the coefficient of the
square term should be $-1$, (ii) the coefficient of the linear
term should be zero, and (iii) the constant term should be
$\ln(\Delta J_{\rm bin}/\sqrt{2\pi}) \approx -0.9$, where $\Delta
J_{\rm bin}$ is the bin-size used to obtain the histogram and is
equal to $0.1\:\rm{second}^{-1}$, i.e., $1$ particle per unit
time. Consistent with these predictions, the coefficient observed
for the square term is $-0.98$, for the linear term is $-0.0018$,
and for the constant term is $-0.94$. The coefficient of
determination for the quadratic fit is $R^2=0.98$. Next, we
analyze the bad actors -- the backward flows -- in two different
ways.

\subsection{The Bad-Actor Trajectory Counts are Well Predicted by the Model.}
\begin{figure}
\centering
\includegraphics*[width = 8cm]{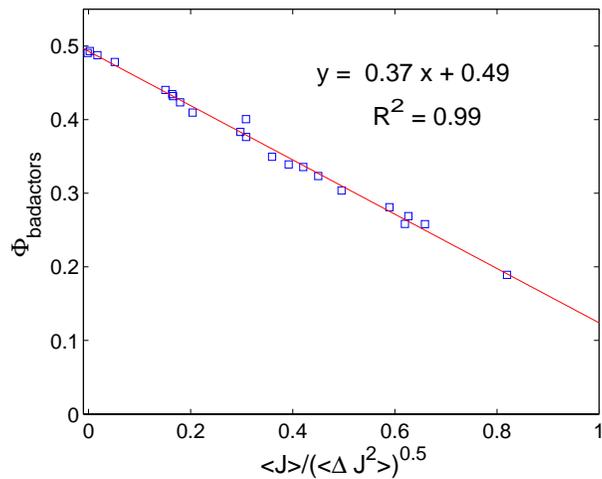}
\caption{\label{fig:badactors} The fraction of trajectories that
are bad-actors vs. the deviation from equilibrium as characterized
by the normalized mean flux, $\la J \ra/\sqrt{\la(\Delta
J)^2\ra}$. Experimental data is shown in squares, while the solid
line represents the fit to the data. The coefficient of
determination ($R^2$) for the fit is also reported. The slope and
intercept agree well with the model.}
\end{figure}

Equation~\ref{eq:badactors} predicts that, for small values of
$\la J \ra/\sqrt{\la(\Delta J)^2\ra}$, the fraction of bad-actors
should be linearly proportional to $\la J \ra/\sqrt{\la(\Delta
J)^2\ra}$. In good agreement, Figure ~\ref{fig:badactors} confirms
this linearity and gives the predicted intercept of $0.5$. This
means that as the system approaches equilibrium (i.e. $\la J \ra
\approx 0$), about half the trajectories involve flow down the
vanishingly small gradient and half the trajectories involve flow
up that small gradient. In the linear regime, the best fit line
shows the slope to be $0.37$, which agrees well with the expected
value of $1/\sqrt{2\pi} \approx 0.4$ from Eq.~\ref{eq:badactors}.
The coefficient of determination for the linear fit is $R^2=0.99$.
Another key feature of this graph is that when the system is
farther away from equilibrium (as implied by a larger mean flux),
the bad-actor fraction reduces. What this means is that more of
the microtrajectories available to the system in this case are
potent to change the current state of the system. Indeed, in our
earlier paper we characterized this idea quantitatively in the
form of the potency~\cite{ghosh06}.

\subsection{The Experiments Confirm the Flux Fluctuation Theorem.}
\begin{figure}
\centering
\includegraphics*[width = 8 cm]{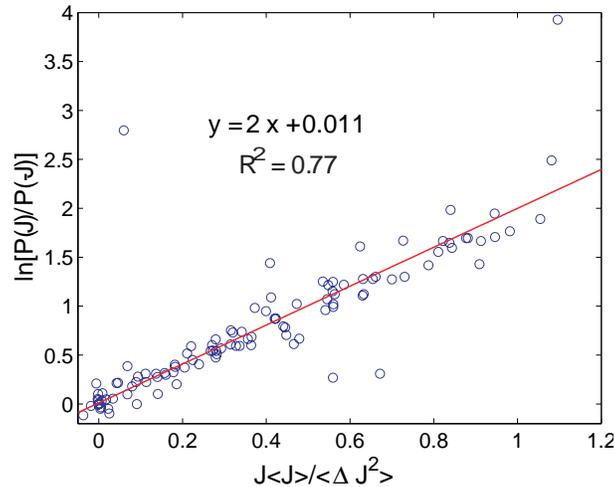}
\caption{\label{fig:masterslope} The flux fluctuation theorem. The
plot shows $\ln [P(J)/P(-J)]$ vs. $\la J \ra J/\la (\Delta J)^2
\ra$ for different values of $\la J \ra$ and $\la (\Delta J)^2
\ra$ arising due to different combinations of $N_1$ and $N_2$.
Experimental data is shown in circles, while the solid line
represents the fit to the data. The coefficient of determination
($R^2$) for the fit is also reported. The slope and intercept
agree with the prediction of Equation~\ref{eq:DF_fluctthm}.}
\end{figure}

Figure~\ref{fig:masterslope} shows $\ln (P(J)/P(-J))$ vs. the
flux, normalized by $\la J \ra J/ \la (\Delta J)^2 \ra$, to
account for different averages and variances of the flux
distribution. This rescaling leads to a linear master curve as
predicted by Eq.~\ref{eq:DF_fluctthm}. Experiments show the slope
to be $2.0$, in perfect agreement with the predicted slope of $2$
from Eq.~\ref{eq:DF_fluctthm}. In this figure, there are a few
aberrant data points that we are unable to explain. However, the
coefficient of determination is still close to unity, $R^2=0.77$.
Though ultimately the fluctuation theorem is a reflection of the
nature of the flux distribution function, such theorems are a
compact way to quantitatively illustrate the significance of
bad-actor microtrajectories.

\subsection{Fick's Law Holds Even in the Small-Numbers
Limit.}
\begin{figure}
\centering \vspace*{0.5cm}
\centering
\includegraphics*[width = 8cm]{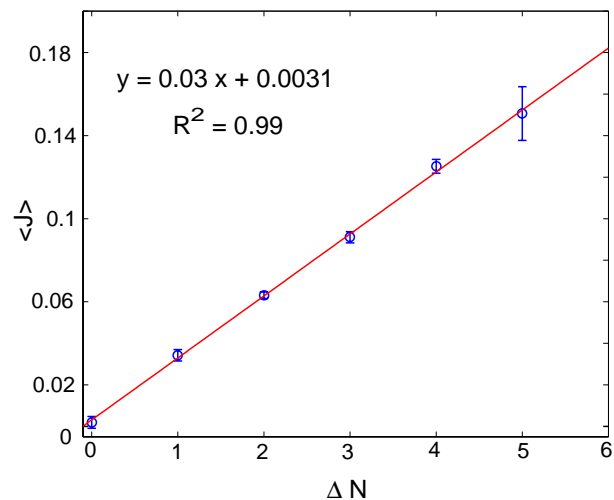}
\caption{\label{fig:JvsDN} Experimental support for Fick's Law,
even down to few-particle gradients. The average flux, $\la J
\ra$, is shown as a function of $\Delta N$, the gradient in the
particle number between two neighboring bins. Experimental data is
shown in circles, while the solid line represents the fit to the
data. The coefficient of determination ($R^2$) for the fit is also
reported. The error bars shown are the variances due to the
different combinations of $N_1$ and $N_2$ resulting in the same
$\Delta N$. The slope and intercept are in agreement with the
expected theoretical values, based on Equation~\ref{eq:J}. }
\end{figure}

We compare the average flux between two neighboring bins, $\la J
\ra$, with the difference in particle numbers, $\Delta N = N_1 -
N_2$. This data is compiled from all the values of $N_1$ and $N_2$
that provide a given $\Delta N$. Fig.~\ref{fig:JvsDN} shows that
$\la J \ra$ depends linearly on the particle number gradient
$\Delta N$, even down to ``gradients'' of zero or one particle,
indicating that Fick's Law holds in the small-numbers limit. The
slope of the graph ($0.03/\rm{second}$) also gives us a value of
the jump rate, $p=0.3$, which is in good agreement with the
theoretical estimate of $0.33$ made in terms of the bin-size and
the diffusion coefficient~\cite{refDD}. As expected, the intercept
is close to $0$ (see Equation~\ref{eq:J}). The linear fit has a
coefficient of determination $R^2=0.99$.

\subsection{The Second Moment of Particle Flux is Proportional to the Sum of Particle Numbers.}
\begin{figure}
\centering \vspace*{0.5cm}
\centering
\includegraphics*[width = 8cm]{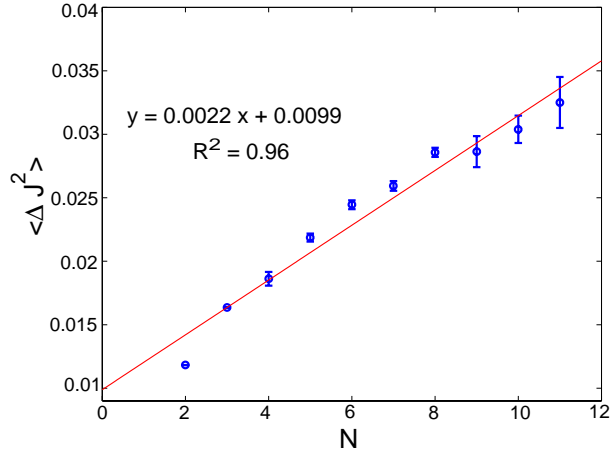}
\caption{\label{fig:dJsqr} The second cumulant, $\la \Delta J^2
\ra = \la J^2\ra - \la J \ra^2$, vs. the total number of
particles, $N$. The second moment of particle flux is proportional
to the sum of particle numbers in the two bins. Experimental data
is shown in circles, while the solid line represents the fit to
the data. The coefficient of determination ($R^2$) for the fit is
also reported. The error bars show the variances due to the
different combinations of $N_1$ and $N_2$ that result in the same
$N$. The slope and intercept are well predicted by
Equation~\ref{eq:DJsqr}. }
\end{figure}

Equation~\ref{eq:DJsqr} predicts that the second moment of the
flux should be proportional to the sum of particle numbers in the
two bins, $N = N_1 + N_2$. Figure~\ref{fig:dJsqr} confirms this
dependence of $\la \Delta J^2 \ra = \la (J - \la J \ra)^2 \ra$ on
$N$. For the optimized particle concentrations, the slope
($0.0022/\rm{second}^2$) is equal to the expected slope of
$0.0022/\rm{second}^{2}$ for the value of $p=0.33$. The
coefficient of determination is $R^2=0.96$. At higher particle
concentrations (data not shown), however, not surprisingly,
systematic errors begin to appear and the slope deviation is quite
high compared to the expected value. We performed Brownian
dynamics simulations that show the likely cause of these
concentration-dependent errors is non-conservation of bin counts,
from particles that either overlap or go out of focus in one
snapshot and into focus in the next (see previous section).

\section{Conclusions}
In a microfluidics experiment, we have determined the distribution
of particle fluxes in few-particle diffusion.  Random-flight and
maximum-caliber models predict a gaussian distribution of fluxes.
Moreover, with a single parameter $p$, which is essentially the
diffusion constant, we find agreement of the theory with several
experimental properties which are usually not examined in
diffusion, and are in direct analogy with quantities of recent
interest in other nonequilibrium experiments~\cite{evans02,
seifert06, bustamante, feitosa04}. First, we find that Fick's law
-- the proportionality of average flux to the gradient of average
concentration -- holds even down to concentration gradients as
small as a single particle. Experiments also confirm that the
variance in the flux is proportional to the total number of
particles, $\la J^2 \ra \: \propto \: N_1 + N_2$, with correct
slopes within experimental errors.  In addition, we describe a
new``flux fluctuation theorem'', that is found to be consistent
with the data in predicting an exponentially diminishing number of
variant trajectories, as a function of the deviation from
equilibrium. The model predicts the backwards flows, the bad
actors, which are relatively infrequent situations in which
particles flow up
--- rather than down --- their concentration gradients and shows
that this subset of the overall repertoire of microscopic
trajectories can be characterized quantitatively.
\section{Acknowledgements}
We thank S. Blumberg, F. Brown, D. Drabold, P. Grayson, L. Han, C.
Jarzynski, J. Kondev, H. J. Lee, H. Qian, S. Quake, S. Ramaswamy,
U. Seifert, T. Squires, Z.G. Wang, E. Weeks, and D. Weitz for
helpful and stimulating comments and discussions. K.A.D. and
M.M.I. would like to acknowledge support from NIH Grant No.
R01GM034993, E.S. acknowledges support from the Betty and Gordon
Moore Fellowship, and R.P. acknowledges support from NSF Grant No.
CMS-0301657, CIMMS, the Keck Foundation, NSF NIRT Grant No.
CMS-0404031, and NIH Director's Pioneer Award Grant No. DP1
OD000217.

\newpage


\begin{references}
\bibitem{refAA}{The
bead concentration is such that the mean particle spacing is about
$8\rm{\mu m}$. For such large particle separations,
electrostatic~\cite{crocker94} and hydrodynamic
interactions~\cite{meiners99, dufresne00} are negligible. Although
the polystyrene particles have a small negative surface charge, it
is readily shown using the Poisson-Boltzmann equation that there
is little interaction at these separations, and we confirmed that
result by an experiment showing an independence on salt
concentrations (data not shown). Also, the data of Meiners and
Quake~\cite{meiners99} and Dufresne~\etal~\cite{dufresne00},
applied to our system, show that hydrodynamic interactions should
be small.}
\bibitem{refCC}{The results obtained under the equilibrium setting were similar to the ones presented in this paper. However, the equilibrium measurements were not presented since the local concentration gradient was obviously rather small, thus limiting the
range of $\Delta N$ values.}
\bibitem{refDD}{The jump probability $p$ in the time interval $\Delta t$ from bin $i-1$ to bin $i$ (of size $\Delta x$) can be
estimated from the diffusion of a particle with diffusion
coefficient $D$. However, there is also a possibility of a jump
occurring from any of the other bins to bin $i$. After summing all
the jump probabilities from all the neighboring bins, the
effective probability $p$ is found to be $p\simeq0.33$, for the
given values of $\Delta x$ and $\Delta t$ and $D$.}
\end{references}
\end{document}